\documentclass[sigconf,natbib]{acmart}
\AtBeginDocument{%
  }

\setcopyright{acmlicensed}
\copyrightyear{2025}
\acmYear{2025}
\acmDOI{XXXXXXX.XXXXXXX}

\usepackage{cleveref}
\usepackage{algorithm}
\usepackage{algpseudocode}
\usepackage{amsmath}

\newcommand{\argmin}{\text{arg\,min}}

\usepackage{dsfont}

\usepackage{interval}
\usepackage{subfiles}
\usepackage{subcaption}
\usepackage{graphicx}
\usepackage{booktabs}

\usepackage{xspace}
\usepackage{todonotes}
\usepackage{soul}

\newcommand{\tuple}[1]{\ensuremath{\langle#1\rangle}\xspace}
\newcommand{\set}[1]{\ensuremath{\{#1\}}\xspace}
\newcommand{\Graph}{\mathcal{G}\xspace}
\newcommand{\card}[1]{\ensuremath{|#1|}\xspace}

\everypar\expandafter{\the\everypar\loosness=-1}
\begin{document}

\title{Demystifying Sequential Recommendations: Counterfactual Explanations via Genetic Algorithms} 


\author{Domiziano Scarcelli}
\email{scarcelli.1872664@studenti.uniroma1.it}
\affiliation{
        \institution{Sapienza University of Rome}
        \city{Rome}
        \country{Italy}
        }
\authornote{These authors contributed equally to this research.} 

\author{Filippo Betello}
    \email{betello@diag.uniroma1.it}
    \affiliation{
        \institution{Sapienza University of Rome}
        \city{Rome}
        \country{Italy}
        }
    \authornotemark[1]

\author{Giuseppe Perelli}
\email{perelli@di.uniroma1.it}
\affiliation{
        \institution{Sapienza University of Rome}
        \city{Rome}
        \country{Italy}
        }

\author{Fabrizio Silvestri}
    \email{fsilvestri@diag.uniroma1.it}
    \affiliation{
        \institution{Sapienza University of Rome}
        \city{Rome}
        \country{Italy}
        }

\author{Gabriele Tolomei}
\email{tolomei@di.uniroma1.it}
   \affiliation{
        \institution{Sapienza University of Rome}
        \city{Rome}
        \country{Italy}
        }


\begin{abstract}
Sequential Recommender Systems (SRSs) have demonstrated remarkable effectiveness in capturing users' evolving preferences. However, their inherent complexity as ``black box'' models poses significant challenges for explainability. This work presents the first counterfactual explanation technique specifically developed for SRSs, introducing a novel approach in this space, addressing the key question: \textit{What minimal changes in a user’s interaction history would lead to different recommendations?} To achieve this, we introduce a specialized genetic algorithm tailored for discrete sequences and show that generating counterfactual explanations for sequential data is an NP-Complete problem. We evaluate these approaches across four experimental settings, varying between targeted/untargeted and categorized/uncategorized scenarios, to comprehensively assess their capability in generating meaningful explanations. Using three different datasets and three models, we are able to demonstrate that our methods successfully generate interpretable counterfactual explanation while maintaining model fidelity close to one. Our findings contribute to the growing field of Explainable AI by providing a framework for understanding sequential recommendation decisions through the lens of ``what-if'' scenarios, ultimately enhancing user trust and system transparency. 

\end{abstract}

\ccsdesc[500]{Information systems~Recommender systems}
\ccsdesc[500]{Computing methodologies~Neural networks}
\ccsdesc[300]{Human-centered computing~Human computer interaction (HCI)}

\keywords{Sequential Recommender Systems, Explainability, Counterfactual Explanations}


\maketitle

\section{Introduction}

Recommender systems (RSs) play a crucial role in everyday life, helping users to navigate through the overwhelming amount of information available online \cite{roySystematicReviewResearch2022}. As e-commerce, social media and streaming platforms continue to grow globally, these systems have become indispensable for tailoring content to individual preferences and enhancing user interaction \cite{zhang2019deep}. RSs analyze vast amounts of user data, processing past interactions and preferences while merging them with contextual information to predict and suggest items that align with individual tastes and behaviors \cite{bobadilla2013recommender}.

In recent years, Sequential Recommender Systems (SRSs) have gained significant popularity as an effective method for modeling user behavior over time \cite{quadrana2018sequence,betellorepro}. By leveraging the temporal dependencies within users’ interaction sequences, these systems can make more precise predictions about user preferences \cite{wangSequentialRecommenderSystems2019}. 

\begin{figure}[t!]     
\centering     
\includegraphics[width=\columnwidth]{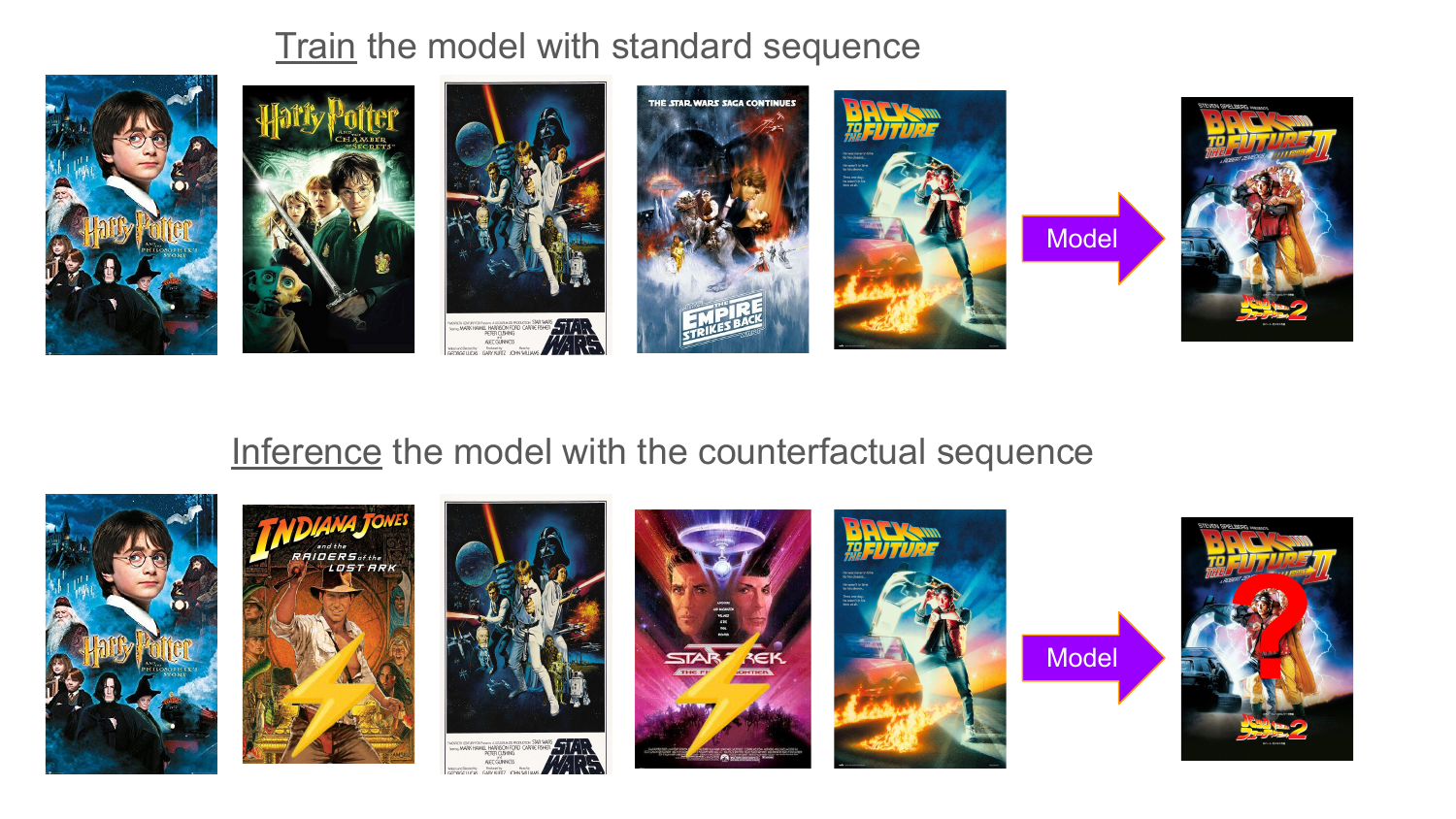}
\caption{Graphical example of our method: We first train the model using the standard sequence and then generate a corresponding counterfactual sequence using the model at inference time.}
\label{fig:GA}
\Description[<Graphical example>]{<Graphical example of our method: We first train the model using the standard sequence and then generate a corresponding counterfactual sequence using the model at inference time.>}
\end{figure}

Despite their widespread adoption and effectiveness, a major challenge remains: the lack of interpretability. Many state-of-the-art recommender systems, particularly those using deep learning and complex machine learning models, operate as ``black boxes'', producing recommendations based on intricate, high-dimensional representations \cite{zhang2020explainable}. The opacity of these models makes it challenging to understand the rationale behind specific recommendations, leading to concerns among users and system developers \cite{zangerle2022evaluating}. Moreover, regulations such as the General Data Protection Regulation (GDPR)\footnote{As outlined in Recital 71: \url{https://gdpr-info.eu/recitals/no-71/}} and Artificial Intelligence Act \cite{kop2021eu} emphasize the right to explanation in automated decision-making processes \cite{voigt2017eu}.


Recently, \textit{counterfactual explanations} (CF) \cite{wachter2017counterfactual} have emerged as a key tool to attach motivation behind items recommended to users. In concrete terms, CFs generate a minimal set of meaningful interactions, without which the recommended items will not end up in the list of suggestions for specific users. 

While extensive research has explored counterfactual explanations for general machine learning models, e.g. \cite{tolomei2017interpretable,montavon2018methods,sharmaCERTIFAICounterfactualExplanations2020}, relatively few studies have adapted these methods for SRSs. For example, ACCENT \cite{tranCounterfactualExplanationsNeural2021} applies counterfactual explanations to neural recommenders, while CauseRec \cite{zhangCauseRecCounterfactualUser2021} leverages counterfactual reasoning for data augmentation in sequential recommendations. However, these methods focus on different applications and do not specifically address counterfactual explanations in the context of SRSs.

To bridge this gap, our work focuses on enhancing the explainability of SRSs by leveraging counterfactual explanations to provide actionable insights into model behavior. By generating such explanations, we seek to offer users a clearer understanding of why a particular item was suggested and what factors influenced the model’s decision. \Cref{fig:GA} illustrates our approach: we first train the model on the standard user sequence and then, at inference time, introduce minimal modifications to the sequence to generate a different recommendation.

We first demonstrate that generating counterfactual explanations for sequential data is an NP-Complete problem. Then, we propose a novel counterfactual generation technique specifically designed for SRSs: \textbf{GECE} (\textbf{GE}netic \textbf{C}ounterfactual \textbf{E}xplanations). This uses a genetic algorithm tailored for discrete sequences, optimizing counterfactual discovery based on sequence similarity and interpretability constraints. By integrating these methods, we aim to improve interpretability in SRSs while maintaining recommendation quality and user trust. We conduct experiments across four distinct contexts, varying in whether counterfactual recommendations are targeted at a specific item and whether they consider item categories rather than individual items. This allows us to assess the impact of semantic constraints and solution sparsity on counterfactual generation.

Our study investigates the utility of counterfactual explanations in SRSs by addressing the following research questions:
\begin{itemize}
\item \textbf{RQ1:} Why is generating counterfactual explanations for sequential data computationally challenging?
\item \textbf{RQ2:} How does enforcing counterfactual constraints on a specific target item impact recommendation outcomes?
\item \textbf{RQ3:} How does generating counterfactuals at the category level, rather than the item level, influence interpretability and effectiveness?
\end{itemize}

We validate our methods using three benchmark datasets and three SRSs models, employing model fidelity as the evaluation metric and Hamming distance as the distance metric. Our findings demonstrate the effectiveness of our proposed techniques in enhancing the interpretability of sequential recommendations while maintaining robust predictive performance. The results provide valuable insights into the potential of counterfactual explanations as a means of increasing transparency and trust in recommender systems.

\section{Related work} \label{sec:related}
In this section, we review existing research on SRSs and counterfactual examples, and conclude with our contributions and distinguishing factors from previous work.

\subsection{Sequential Recommender Systems}
Early recommender systems relied on Markov chain models \cite{fouss2005web}, which were simple and efficient but struggled to capture long-term dependencies in sequential data. With the advent of deep learning, recurrent neural networks (RNNs) emerged as a powerful alternative \cite{donkers2017sequential,hidasi2016sessionbased}. These models encode a user's historical preferences into a continuously updated vector, enabling sequential prediction. However, RNNs often face challenges in maintaining long-term dependencies and generating diverse recommendations.

The introduction of the attention mechanism \cite{vaswani2017attention} marked a significant breakthrough in sequential recommendation. Models such as SASRec \cite{kang2018self} and BERT4Rec \cite{sun2019bert4rec} use attention to dynamically weight different parts of the sequence, allowing them to focus on the most relevant interactions and improve prediction accuracy.

More recently, Graph Neural Networks (GNNs) have gained traction in sequential recommendation \cite{chang2021sequential,fan2021continuous}. By modelling complex relationships between items and user interactions, GNN-based approaches further enhance the ability of sequential recommender systems (SRSs) to capture intricate dependencies \cite{wu2022graph,purificatosheafacm}.

\subsection{Counterfactual Explanations}
Several approaches have been proposed to enhance interpretability in machine learning models. In counterfactual explanations, approaches like FACE \cite{poyiadziFACEFeasibleActionable2020} prioritize feasibility and actionability, while MACE \cite{karimiModelAgnosticCounterfactualExplanations} frames counterfactual search as a Boolean satisfiability problem. Other methods, such as LORE \cite{guidottiLocalRuleBasedExplanations2018} and CERTIFAI \cite{sharmaCERTIFAICounterfactualExplanations2020}, use genetic algorithms to optimize counterfactual generation, balancing similarity to the original input with predictive validity. Other works use Reinforcement Learning \cite{chen2022relax} or Random Forest based models \cite{tolomei2017interpretable} to provide efficient counterfactual explanations.

While the aforementioned methods focus on general counterfactual explanations, few works have been specifically tailored for recommender systems. ACCENT \cite{tranCounterfactualExplanationsNeural2021} introduces a novel approach to counterfactual reasoning in recommendation systems, specifically for neural recommenders: it generates counterfactuals based on a user’s own past interactions. \cite{zhangCauseRecCounterfactualUser2021, wang2021counterfactual} utilize counterfactual examples as a data augmentation technique to address the inherent sparsity of recommender system datasets. \cite{ghazimatin2020prince,kaffes2021model} both investigate perturbations to user interaction histories: however, the former assumes a modified recommendation engine through pruning, while the latter restricts its scope to deletions, whereas our approach encompasses a broader range of perturbation types.  \citet{tan2021counterfactual} and \citet{ranjbar2024explaining} examine perturbations to item features, while \citet{chen2023dark} uses counterfactuals to poison a recommender system.

To the best of our knowledge, this is the first work to leverage counterfactual examples to enhance explainability in the field of SRSs.

\section{Method}
In this section, we present the proposed method and its formulation. We begin by providing the necessary background and preliminaries, followed by an analysis of the complexity involved in finding a counterfactual for a sequence. Finally, we introduce our approach for generating counterfactual sequences.

\subsection{Background and Preliminaries} \label{sec:backgroundandprel}
The primary goal of an SRS is to predict the next interaction within a sequence. Let $\mathcal{U} \subset \mathbb{N^+}$ be a collection of $n$ users and $\mathcal{I} \subset \mathbb{N^+}$ a set of $m$ items. Each user $u$ is represented by a temporally ordered sequence of interactions $S_u = (i_1, i_2, \dots, I_{L_u})$ with which it has interacted, where $L_u$ is the length of the sequence and $i_j \in \mathcal{I}$. An SRS $\mathcal{M}$ takes as input the user's sequence up to the $l$-th item, $S_{u_l} = [s_1, \dots, s_l]$, and aims to predict the next item, $s_{l+1}$.

We recall that counterfactual explanations stand out as one of the most powerful post-hoc tools for understanding predictions of individual instances in the form: ``{\em If}  A {\em had been different,} B {\em would not have occurred}''. At the core of counterfactual explanations is the search for so-called counterfactual examples, i.e., modified versions of input samples that result in alternative output responses from the predictive model. Typically, the problem of generating counterfactual examples is formulated as an optimization task, whose goal is to find the ``closest'' data point to a given instance, which crosses the decision boundary induced by a trained predictive model. Concretely, we make the following assumptions:
\begin{itemize}
\item[$(i)$] The length of the original ($S_{u_l}$) and perturbed ($S_{u_l}'$) sequence must be the same ($L_u$).
\item[$(ii)$] Every element of the original sequence can be replaced with any other element of the set of items $\mathcal{I}$. If we assume $|\mathcal{I}|=n$, the space of possible counterfactual sequence candidates is $P(n,L_u) = \frac{n!}{(n-L_u)!}$. Notice that this also covers all the $L_u$! possible rearrangements of the items of the input sequence.
\end{itemize}

\begin{definition}[Valid Counterfactual Sequence]
\label{def:valid-cs}
Let $\mathcal{I}$ be a finite set of items and $S_{u_l}\in \mathcal{I}$ a sequence of items of length $L_u > 0$. 
Additionally, let $\mathcal{M}(S_{u_l})$ be the output of a predictive model when input with the sequence $S_{u_l}$.
A valid counterfactual sequence for $S_{u_l}$, if it exists, is another $L_u$-length sequence $S_{u_l}'\neq S_{u_l}$, such that $\mathcal{M}(S_{u_l}')\neq \mathcal{M}(S_{u_l})$.
\end{definition}

Given an input sequence $S_{u_l}$, there can be zero or more counterfactual sequences associated with it with respect to a specific model $\mathcal{M}$. 
If we assume that at least one counterfactual sequence exists for $S_{u_l}$, we can  measure the ``cost'' of transforming $S_{u_l}$ in to each counterfactual sequence $S_{u_l}', S_{u_l}'', \ldots$. 
Such a cost must take into account the effort needed to change the original sequence into the counterfactual one. 
Generally speaking, we consider a distance function $\delta: I^* \times I^* \mapsto \mathbb{R}_{\geq 0}$ that takes as input the original and the counterfactual sequences and produces as output a positive real value. For example, $\delta$ can be defined as the edit (Hamming) distance or a smoothed differentiable variant of it \cite{ofitserov2019soft}.

\begin{definition}[Optimal Counterfactual Sequence]
\label{def:opt-cs}
Let $S_{u_l}\in \mathcal{I}^k$ be a sequence of symbols of length $L_u > 0$. 
Let $\mathcal{C}_{S_{u_l}}$ be the union set of all the valid counterfactual sequences for $S_{u_l}$. 
Furthermore, let $\delta: I^*\times I^* \mapsto \mathbb{R}_{\geq 0}$ be a function that measures the distance between two sequences.
The optimal counterfactual sequence $S_{u_l}^*$ for $S_{u_l}$, if it exists (i.e., if $\mathcal{C}_{S_{u_l}} \neq \emptyset$), is defined as follows:
\[
{S_{u_l}}^* = \argmin_{S_{u_l}'\in \mathcal{C}_{S_{u_l}}}\{\delta({S_{u_l}}',{S_{u_l}})\}.
\]
\end{definition}

\subsection{Problem Complexity} \label{sec:problem_complexity}

We consider the problem of finding the optimal counterfactual sequence as defined in Def.~\ref{def:opt-cs} above.

First of all, notice that a straightforward solution to this problem is given by a trivial algorithm that exhaustively enumerates all the possible modifications of the original input sequence. 
For each valid counterfactual sequence found, this solution keeps track of the one ``closest'' to the original input sequence (according to $\delta$).
Eventually, it returns either the optimal counterfactual sequence or $\perp$ if none exists.
Clearly, this algorithm works, but it may be computationally unfeasible due to the extremely large space of solutions to explore. Indeed, for a $L_u$-length input sequence from a set of $|\mathcal{I}|$ items, this solution must enumerate all the possible $L_u$-permutations $P(n,L_u)=\frac{n!}{(n-L_u)!}$.

Of course, this also depends on the internal workings of the model $\mathcal{M}$. For instance, if $\mathcal{M}$ primarily relies on the last element of the input sequence to generate its output, then altering the last element of the sequence might result in a different output. Consequently, this change could yield a valid counterfactual sequence.
However, in general, we cannot make any assumptions about the inner logic of the model. As a matter of fact, given its typical complexity, the model must be treated as a ``black box.''

To better analyze this problem, let us consider its corresponding decision problem. This involves determining whether there exists an $\varepsilon$-valid counterfactual sequence $S_{u_l}'$ for $S_{u_l}$, such that the distance between $S_{u_l}$ and $S_{u_l}'$ is no greater than a small, positive constant $\varepsilon$.

\begin{definition}[$\varepsilon$-Valid Counterfactual Sequence {[}$\varepsilon$-VCS{]}]
\label{def:eps-cs}
Let $S_{u_l}'$ be a valid counterfactual sequence for $S_{u_l} \in \mathcal{I}$, and $\varepsilon \in \mathbb{R}_{>0}$ a positive real-valued constant.
Therefore, $S_{u_l}'$ is an $\varepsilon$-valid counterfactual sequence for $S_{u_l}$ iff $\delta(S_{u_l}',S_{u_l}) \leq \varepsilon$.
\end{definition}

As it turns out, deciding whether there exists an $\varepsilon$-valid counterfactual sequence for a given input sequence is an NP problem. In other words, the decision version of the problem of finding the optimal counterfactual sequence for a given input sequence is in NP.

\begin{lemma}
\label{lemma:np}
Deciding $\varepsilon$-VCS is in NP
\end{lemma}
\begin{proof}
Let $\widetilde{S_{u_l}}$ be a certificate for a possible $\varepsilon$-valid counterfactual sequence for the input sequence $h$. If we assume the model $\mathcal{M}$ and the computation of the distance $\delta$ can both run in polynomial time, we can build a verifier that also runs in polynomial time (since the composition of polynomial functions is still a polynomial) as follows:
\begin{algorithm}[H]
\small
\begin{algorithmic}[1]
\If {$\mathcal{M}(\widetilde{S_{u_l}}) == \mathcal{M}(S_{u_l})$}
\State \textbf{return false} \Comment $\widetilde{S_{u_l}}$ is not a valid counterfactual sequence;
\EndIf
\If {$\delta(\widetilde{S_{u_l}},S_{u_l}) > \varepsilon$}
\State \textbf{return false} \Comment $\widetilde{S_{u_l}}$ is too far from $S_{u_l}$;
\EndIf
\State \textbf{return true} \Comment the certificate is valid;
\end{algorithmic}
\end{algorithm}
\end{proof}

So far, we have shown that $\varepsilon$-VCS is in NP. Furthermore, we show that it is also NP-hard and, therefore, NP-complete.

\begin{theorem}
	\label{thm:np}
	Deciding $\varepsilon$-VCS is NP-complete.
\end{theorem}
\begin{proof}
We have already shown that $\varepsilon$-VCS is in NP.
Therefore, we must show that $\varepsilon$-VCS is NP-hard.
To do so, we must find a polynomial-time reduction from a well-known NP-complete problem.
Below, we demonstrate that such a polynomial reduction exists from \textsc{VertexCover}.

Consider an instance for Vertex Cover, which is an undirected graph $\Graph = \tuple{V, E}$, with $V = \set{v_{1}, v_{2}, \ldots, v_{n}}$ and $E \subseteq V \times V$.
Then define the set of items $\mathcal{I} = \set{v_{1}, \overline{v_{1}} v_{2}, \overline{v_{2}} \ldots, v_{n}, \overline{v_{n}}, \bot, \top}$ given by \emph{positive} and \emph{negative} instantiation of the vertexes in $V$, together with two special symbols $\bot$ and $\top$.
Now, define the SRS $\mathcal{M}$ over the set of items $\mathcal{I}$ defined above as, such that, for every sequence of the form $S = [\iota_1, \ldots \iota_n]$ with $\iota_i \in \set{v_i, \overline{v_i}}$ being either a positive or negative occurrence of vertex $v_i$, is such that

\begin{center}
	$\mathcal{M}(S) =
	\begin{cases}
		\top & \text{ if } S \cap V \text{ is a VC of } \Graph \\
		\bot & \text{ otherwise }
	\end{cases}$.
\end{center}

Intuitively, the SRS is mimicking the verification that a given subset of $V$, identified in $S$ with its positive occurrences of vertexes, is a vertex cover of $\Graph$.

Now, assuming $\delta$ to be the edit distance, we prove that $\Graph$ has a vertex cover of size at most $k$ if, and only if, the sequence $\overline{S} = [\overline{v_1}, \ldots \overline{v_{n}}]$ admits a $k$-valid counterfactual sequence.

From left to right direction, assume $\Graph$ has a vertex cover $C \subseteq V$ of size at most $k$, that is $\card{C} \leq k$ where

\begin{center}
	$\iota_i = \begin{cases}
		v_i & v_i \in C \\
		\overline{v_i} & v_i \notin C
	\end{cases}$.
\end{center}

That is, $S_C$ is the sequence identifying the cover $C$ by making its and only vertexes occur positively.
Clearly, by construction, it holds that $\mathcal{M}(S_C) = \top \neq \bot = \mathcal{M}(\overline{S})$.
Moreover, $\delta(S_C, \overline{S}) = \card{C} \leq k$.
Therefore $S_C$ is a $k$-valid counterfactual sequence of $\overline{S}$.

From right to left direction, assume that $S'$ is a $k$-valid counterfactual sequence of $\overline{S}$.
Therefore, we must have $\mathcal{S'} = \top$.
This implies, by construction of $\mathcal{M}$ that, $S' = [\iota_1, \ldots, \iota_n]$ with $\iota_i \in \set{v_i, \overline{v_i}}$, for every $i \leq n$.
In particular, $S'$ identifies the subset $C_{S'} \subseteq V$ obtained by including all and only the positive occurrences of vertexes in $S'$.
Furthermore, observe that, following from the fact that $\mathcal{S'} = \top$, it must follow that $C_{S'}$ is a vertex cover of $\Graph$.
In addition, notice that $\card{C_{S'}} = \delta(S', \overline{S}) \leq k$.
Hence there exists a vertex cover of $\Graph$ of size at most $k$.
\end{proof}

We have demonstrated that the problem of finding a counterfactual for sequences is NP-Complete answering to our first research question, making it highly challenging to address. In the next subsection, we introduce the algorithm we have chosen to tackle this task.

\subsection{GECE}

\begin{figure}[b!]     
\centering     
\includegraphics[width=1\columnwidth]{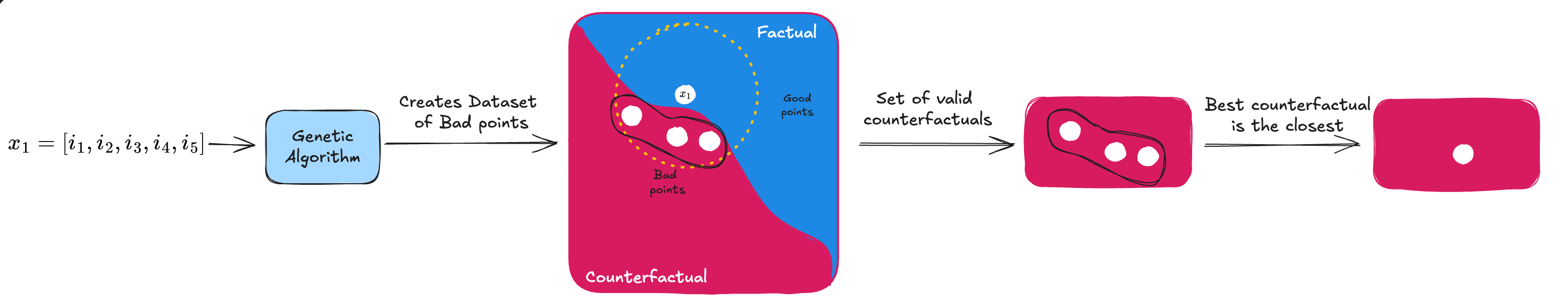}
\caption{Diagram showing GECE's pipeline, generating a counterfactual starting from the original sequence.}
\label{fig:gene}
\end{figure}

GECE (\textbf{GE}netic \textbf{C}ounterfactual \textbf{E}xplanations) is a novel counterfactual explanation framework tailored for sequential recommender systems (\Cref{fig:gene}). It extends the LORE framework \cite{guidottiLocalRuleBasedExplanations2018}, which employs a genetic algorithm, to accommodate ordered sequences of discrete items. The central objective of GECE is to generate counterfactual sequences efficiently by leveraging genetic algorithms to explore the space of possible sequences that best satisfy counterfactual conditions. Genetic algorithms, inspired by the principles of natural selection, operate as metaheuristics for optimization, iteratively refining a population by assessing fitness, applying mutation and crossover, and selecting the fittest individuals until convergence \cite{kramer2017genetic}.

GECE iteratively modifies a given source sequence through a set of mutation operations, optimizing toward a predefined counterfactual objective. Unlike LORE, which is designed for tabular data, GECE is specifically adapted for ordered sequences of discrete, non-repeating items, making it particularly suitable for sequential recommendation tasks. The main adaptation required for sequential data is the mutation operations, as selection and crossover inherently function with item sequences and remain unchanged from LORE.

Given a sequence of items $S_{u_l}$ and a set of candidate items $\mathcal{I}$, GECE defines three mutation operations suited for ordered sequences of discrete, non-repeating items:

\begin{enumerate}
\item
\textit{Replace}: Randomly select an index $i$ within the bounds of $S_{u_l}$ and a random item $z \in \mathcal{I}$ that is not already present in $S_{u_l}$. Replace the item at index $i$ in $S_{u_l}$ with $z$, i.e., $S_{u_l}[i] = z$.
\item
\textit{Add}: Randomly select an index $i$ within the bounds of $S_{u_l}$ and a random item $z \in \mathcal{I}$ not present in $S_{u_l}$. Insert $z$ at position $i$, shifting all subsequent elements to the right: $S_{u_l}[j+1] = S_{u_l}[j]$ for all $j \geq i$, followed by $S_{u_l}[i] = z$.
\item
\textit{Delete}: Randomly select an index $i$ within the bounds of $S_{u_l}$, remove the item at position $i$, and shift all subsequent elements leftward.
\end{enumerate}

\begin{algorithm}[t]
\caption{\textbf{GECE}: GEnetic Counterfactual Explanations}
\begin{algorithmic}[1] \label{alg:GECE}
\Require Source sequence $source$, setting $setting$, target $target$, recommendation model $model$, top-k value $k$
\Ensure Counterfactual sequence $result$
\State $candidates \gets$ \Call{genetic}{$source, setting, target, model, k$}
\State $result \gets \min\{\, \text{edit\_distance}(source, candidate) : \forall\, candidate \in candidates \,\}$
\State \Return $result$
\end{algorithmic}
\end{algorithm}

\begin{algorithm}[t]
\caption{Genetic Algorithm Subroutine}
\begin{algorithmic}[1]
\Require Source sequence $source$, setting $setting$, target $target$, recommendation model $model$, top-k value $k$
\Ensure Final evolved population
\State Initialize population $population$ with $N$ identical copies of $source$
\State Compute $source\_logits$ as the top-$k$ predictions from $model(source)$
\For{each generation in $number\_generations$}
    \State Apply mutation: randomly modify a subset of the population based on $mutation\_probability$
    \State Apply crossover: combine traits from different individuals based on $crossover\_probability$
    \State Compute fitness scores for each individual:
    \State \hspace{1em} - Distance from source: edit distance between $source$ and the individual
    \State \hspace{1em} - Distance from target: difference between $source$'s logits and the individual's logits
    \State \hspace{1em} - Weighted sum of the two distances determines fitness
    \State Select the best individuals to maintain a constant population size
\EndFor
\State \Return $population$
\end{algorithmic}
\end{algorithm}

The methodology for generating counterfactual sequences via a genetic algorithm is outlined in \Cref{alg:GECE}. The objective is to transform an initial sequence (\textit{source}, $S_{u_l}$) into a similar yet distinct sequence (\textit{target}, $S'_{u_l}$) that satisfies a specific counterfactual condition while preserving the underlying assumptions described in \Cref{sec:backgroundandprel}.

The algorithm initializes a population consisting of identical copies of the source sequence. Each generation applies two genetic operations: \textit{Mutation}, wherein a subset of individuals undergoes random modifications with a given probability; and \textit{Crossover}, wherein traits from different individuals are recombined to create new sequences.
These operations introduce diversity into the population, enhancing the likelihood of identifying a suitable counterfactual.

To guide evolution, candidates are assigned a fitness score based on two criteria: \textit{Edit distance}, quantifying the structural divergence between the candidate and the source sequence; and \textit{Logit similarity} (i.e. model’s internal outputs), capturing the change in model-generated logits between the source and candidate sequences, thereby reflecting variations in recommendation behavior.

The overall fitness is determined via a weighted sum of these components. Following fitness evaluation, a selection process retains the highest-performing individuals, ensuring a stable population size across generations. This iterative process continues for a predefined number of generations, progressively refining the population toward sequences that balance similarity to the source and conformity to the target condition. The optimal counterfactual is ultimately selected as the sequence exhibiting the minimal edit distance from the source among the final population.

By leveraging genetic search, GECE enables the efficient discovery of meaningful counterfactual explanations, facilitating the exploration of alternative recommendation outcomes in a structured and computationally effective manner.

\subsubsection{Baselines}
To the best of our knowledge, no existing methods currently address counterfactual reasoning for sequential recommendation, thus there are no direct competitors. However, to demonstrate the effectiveness of our proposed approach, we implement two baseline methods based on common strategies. In one case, named \textit{random}, we randomly select items and substitute them at arbitrary positions within the sequence. In the second case, named \textit{educated}, used only when targeting the specific items selected in the ``Targeted'' scenario (see \Cref{sec:counterfactual_specifications}), we select those items and replace them at random positions in the sequence. These baselines allow us to highlight the strengths and effectiveness of our method, which is the first to apply counterfactual reasoning to sequential recommendation. We chose not to include any of the methods presented in \Cref{sec:related}, as we believe that adapting techniques developed for different domains to the context of sequential recommendation constitutes a substantial research effort on its own. As such, it falls outside the scope of this work.

\section{Experimental Setup}
In this section, we present the datasets selected for our task, along with the models employed and the evaluation protocol adopted.

\subsection{Datasets}

\begin{table}[t!]
    \caption{Dataset statistics after pre-processing; users and items not having at least 5 interactions are removed. Avg. and Med. refer to the Average and Median of $\frac{\mathrm{Actions}}{\mathrm{User}}$, respectively.}
    \resizebox{\columnwidth}{!}{
      \begin{tabular}{c||ccc|ccc}
        \toprule
        Name & Users & Items & Interactions &  Density & Avg. \\ 
        \midrule
        ML-100k & 943 & 1,349 & 99,287 & 7.805 & 105 \\
        ML-1M &  6,040 & 3,416 & 999,611 & 4.845 & 165\\
        Steam & 334,536 & 13,046 & 4,212,143 & 13 & 8 \\
        \bottomrule
      \end{tabular}}
\label{tab:dataset_info}
\end{table}

We use three different datasets:

\begin{itemize}
    \item MovieLens \cite{10.1145/2827872}: Frequently utilized to evaluate recommender systems, this benchmark dataset is employed using both the 100K and 1M versions. This dataset has 17 categories.
    \item Steam \cite{kang2018self}: This dataset originates from a major online video game distribution platform. It contains user interaction data, including game purchase history, playtime, and reviews. This dataset has 22 categories. 
\end{itemize}

The statistics for all the datasets are shown in \Cref{tab:dataset_info}. We are aware that recently MovieLens dataset has been criticized \cite{fan2024our}, however we are using this dataset to demonstrate the feasibility of our method, and moreover it has been recognized as a sequential dataset \cite{klenitskiy2024does}. Our pre-processing technique adheres to recognised principles, such as treating ratings as implicit, using all interactions without regard to the rating value, and deleting users and things with fewer than 5 interactions \cite{kang2018self, sun2019bert4rec}.
For testing, as in \cite{sun2019bert4rec, kang2018self}, we keep the most recent interaction for each user, while for validation, we keep the second to last action. All the remaining sequence is used for training the models. For our task, we train all models using the standard sequence and subsequently perform inference using the counterfactual sequence.

\subsubsection{Counterfactual Dataset Setting} \label{sec:counterfactual_specifications}
We conduct experiments across four distinct contexts, varying in whether counterfactual recommendations are targeted at a specific item and whether they consider item categories rather than individual items. 
For category-based analysis, we randomly selected five categories—maintaining a similar ratio as the item selection process—due to the large number of available categories. The selected categories are: "Horror", "Action", "Adventure", "Animation", "Fantasy", and "Drama" for MovieLens, and "Action", "Indie", "Free to Play", "Sports", and "Photo Editing" for Steam. These experimental settings are:

\begin{itemize} \item Untargeted-Uncategorized: Generating counterfactual sequences with minimal changes to alter the recommendation, without any constraints on item categories.
\item Untargeted-Categorized: Introducing item categories to assess how counterfactual explanations vary when considering higher-level semantic shifts. 
\item Targeted-Uncategorized: Ensuring that the counterfactual leads to a specific desired recommendation, increasing the challenge of solution sparsity. 
\item Targeted-Categorized: Combining category constraints with a target recommendation, providing structured and meaningful explanations. 
\end{itemize}

\begin{figure*}
     \begin{subfigure}{\linewidth}
        \centering
        \includegraphics[width=\linewidth]{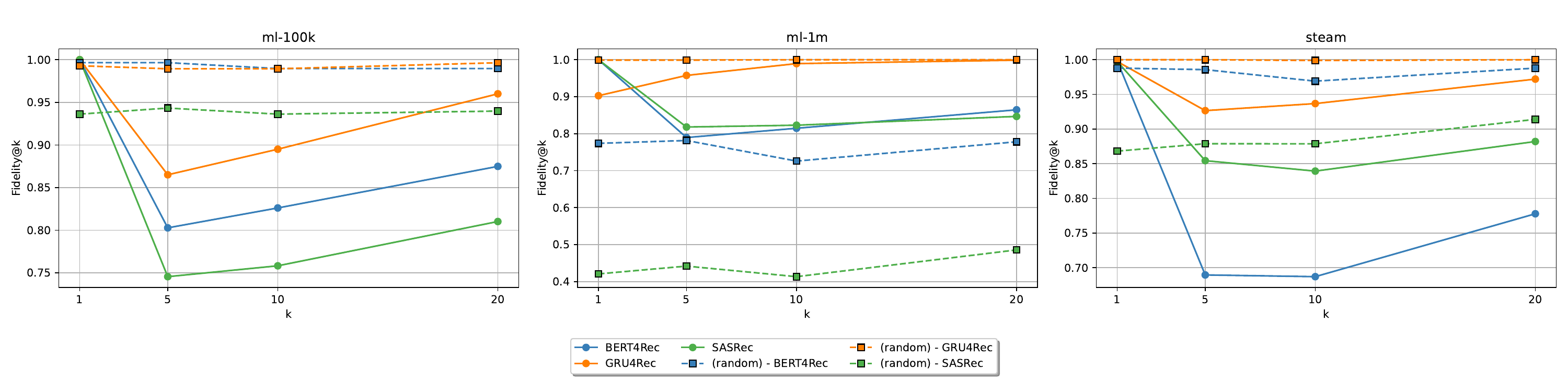}
        \caption{Untargeted and Uncategorized setting} \label{fig:un_un}
    \end{subfigure}
    
    
    \begin{subfigure}{\linewidth}
        \centering
        \includegraphics[width=\linewidth]{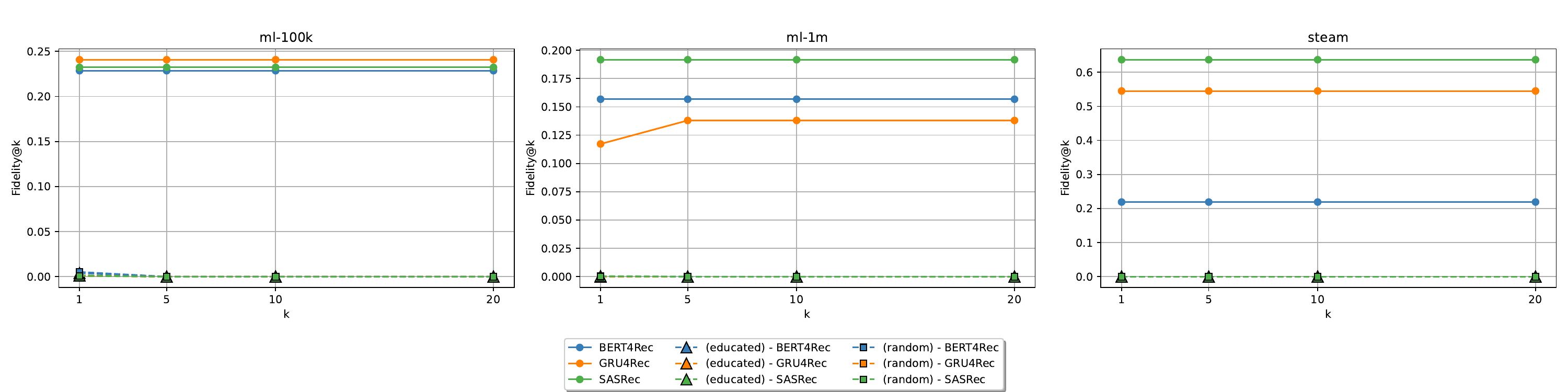}
        \caption{Targeted and Uncategorized setting}\label{fig:tag_un}
    \end{subfigure}
    \caption{Plots of the Fidelity@k in the cases of untargeted and targeted scenarios, always keeping the uncategorized setting. As we can see, the problem is easy in the case of untargeted confirming some previous works while for the targeted scenario becomes difficult due to the cardinality of $\mathcal{I}$.}
    \Description[<Untargeted-Targeted scenarios>]{<Plots of the model fidelity@k in the cases of untargeted and targeted scenarios. As we can see, the problem is easy in the case of untargeted confirming some previous works while for the targeted scenario becomes difficult due to the cardinality of item set $\mathcal{I}$.>}
\end{figure*}

\subsection{Models}
We employ three different models well established in literature:
\begin{itemize}
    \item BERT4Rec \cite{sun2019bert4rec}: This model is based on the BERT architecture, enabling it to capture complex relationships in user behaviour sequences through bidirectional self-attention.
    \item GRU4Rec \cite{hidasi2016sessionbased}: This model utilizes GRUs to capture temporal dependencies in user-item interactions.
    \item SASRec \cite{kang2018selfattentive}: this model is characterized by its use of self-attention mechanisms, allowing it to discern the relevance of each item within the user's sequence.
\end{itemize}

We chose these models because they have demonstrated exceptional performance in numerous benchmarks.
Moreover, since two models are based on attention mechanisms and the other on RNNs, their different network operations make it particularly interesting to evaluate their behaviour.

\subsection{Metrics}

\begin{figure*}
     \begin{subfigure}{\linewidth}
        \centering
        \includegraphics[width=\linewidth]{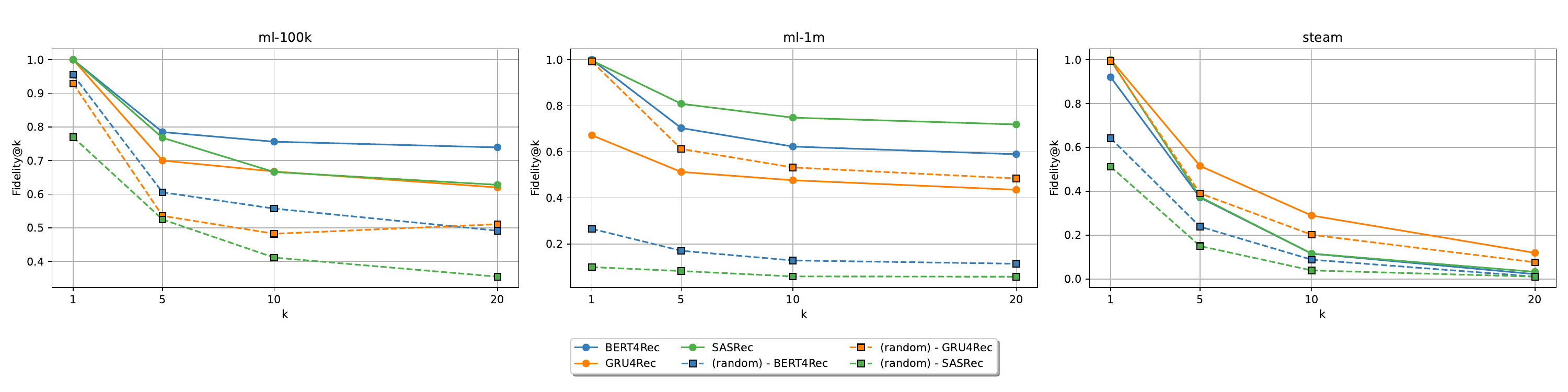}
        \caption{Untargeted and Categorized setting} \label{fig:un_cat}
    \end{subfigure}
    
    
    \begin{subfigure}{\linewidth}
        \centering
        \includegraphics[width=\linewidth]{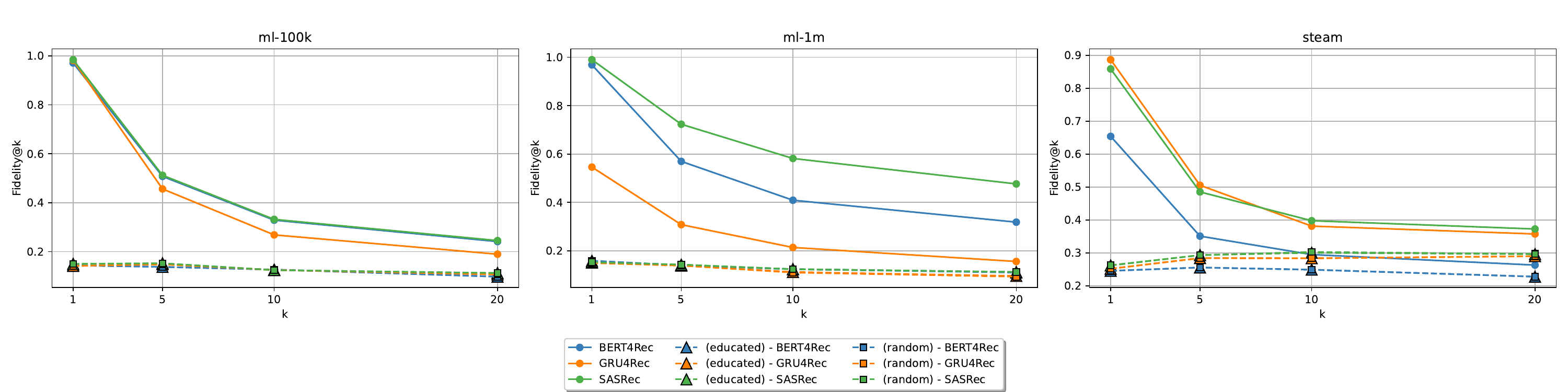}
        \caption{Targeted and Categorized setting}\label{fig:tag_cat}
    \end{subfigure}
    \caption{Plots of the Fidelity@k in the cases of untargeted and targeted scenarios, always keeping the categorized setting. In these scenarios our approach is significantly better than the baselines ($p<0.01$).}
    \Description[<Uncategorized-categorized scenarios>]{<Plots of the model fidelity@k in the cases of untargeted and targeted scenarios. As we can see, the problem is easy in the case of untargeted confirming some previous works while for the targeted scenario becomes difficult due to the cardinality of $\mathcal{I}$.>}
\end{figure*}

Since our task focuses on identifying counterfactuals for sequences rather than ranking items, standard Information Retrieval (IR) metrics such as NDCG or RLS are not suitable. Instead, we adopt evaluation metrics commonly used in explainability, specifically model fidelity and edit distance.

Model Fidelity, a widely used metric \cite{anand2022explainable}, measures the proportion of valid counterfactuals among all generated explanations. Given a top-$k$ extraction and the corresponding set of scores $S_k$, along with threshold $t$, model fidelity is computed as:
\begin{equation*} \text{Fidelity@k} = \frac{1}{|S_k|} \sum_{i=1}^m \mathds{1}(S_k[i] \geq t) \end{equation*}

where $\mathds{1}$ is an indicator function that returns 1 if the condition is satisfied and 0 otherwise. When Fidelity@k equals one, it indicates that all of the top-k explanations meet the validity criterion defined by the threshold $t$. Conversely, a Fidelity@k of zero means that none of the top-k counterfactuals are considered valid.

Additionally, we employ edit distance to quantify the similarity between the generated counterfactual and the original sequence, serving as a measure of the modification required to obtain the counterfactual. Specifically, we use the Hamming distance \cite{norouzi2012hamming}, which counts the number of positions at which the corresponding elements of two sequences of equal length differ.

Given two sequences $S_{u_l} = (i_1, i_2, \dots, i_l)$ and $S'_{u_l} = (i'_1, i'_2, \dots, i'_l)$ of equal length, the Hamming distance $d(S_{u_l}, S'_{u_l})$ is defined as:

\[
d(S_{u_l}, S'_{u_l}) = \sum_{i=1}^{l} \mathds{1}(i_i \neq i'_i)
\]

To evaluate the overall modification across all users, we compute the average Hamming distance:

\[
\text{Mean Hamming Distance} = \frac{1}{N} \sum_{l=1}^{N} d(S_{u_l}, S'_{u_l})
\]

where $N$ is the total number of users, $S_{u_l}$ is the original sequence for user $u_l$, and $S'_{u_l}$ is the corresponding counterfactual sequence.

\subsection{Evaluation}
For our experiments, we utilize the RecBole library \cite{recbole[2.0]} and conduct all computations on a single NVIDIA RTX A6000 GPU with 10752 CUDA cores and 48 GB of RAM. Models are trained for 300 epochs, followed by inference using the hyperparameters specified in their respective original papers. We adopt this approach due to computational constraints and because our primary focus is on the GECE method. Moreover, since the models are only used at inference time, further optimization is unnecessary in this context.

We set the sequence length to 50 to balance computational efficiency with counterfactual search quality, and the batch size to 128. For our genetic algorithm, we perform hyperparameter optimization and fix the paramers to 30 generations, with a population size of 8192, a mutation probability of 50\% and a crossover probability of 70\%. The threshold for a counterfactual to be valid has been set to 0.5.

Target items were randomly sampled from three distinct scenarios—popular, standard, and unpopular—across three different random seeds, and the reported metrics represent the average over these runs.

For the category-based analysis, we chose five categories at random, ensuring the selection ratio mirrored that of the item selection process, given the extensive number of available categories. The selected categories are reported in \Cref{sec:counterfactual_specifications}. Due to space constraints, we report average metric values across these categories using three different random seeds.

We remind the reader that models are trained on the original datasets, and user sequences are modified at inference time using GECE or baseline methods. Specifically, we alter sequences for 943 users in ML-100k, and randomly sample 200 users for ML-1M and Steam, repeated across three seeds, due to computational constraints. Since our goal is to evaluate the effectiveness of the approach rather than exhaustively test the entire user base, we report averaged results over these samples.


\section{Results}

\begin{table*}[t!]
\resizebox{\textwidth}{!}{
\begin{tabular}{c||c|c|c|c||c|c|c|c||c|c|c|c}
\toprule
 & \multicolumn{4}{c||}{Steam} & \multicolumn{4}{c||}{ML-1M} & \multicolumn{4}{c}{ML-100K} \\
\midrule
 & un\_un & targ\_un & un\_cat & targ\_cat & un\_un & targ\_un & un\_cat & targ\_cat & un\_un & targ\_un & un\_cat & targ\_cat \\
 \midrule
SASRec & 1.0 & 1.601 & 1.005 & 1.172 & 1.002 & 1.72 & 1.005 & 1.031 & 1.0 & 1.856 & 1.0 & 1.093 \\
BERT4Rec & 1.013 & 2.352 & 1.201 & 1.483 & 1.001 & 1.586 & 1.012 & 1.088 & 1.0 & 2.05 & 1.001 & 1.108 \\
GRU4Rec & 1.0 & 1.899 & 1.005 & 1.305 & 1.0 & 2.037 & 1.008 & 1.076 & 1.0 & 2.218 & 1.002 & 1.199 \\
\midrule
SASRec (random) & 1.421 & 1.219 & 1.467 & 1.219 & 1.444 & 1.334 & 1.413 & 1.334 & 1.511 & 1.326 & 1.489 & 1.326 \\
BERT4Rec (random) & 1.493 & 1.341 & 1.476 & 1.341 & 1.472 & 1.338 & 1.421 & 1.338 & 1.471 & 1.339 & 1.471 & 1.339 \\
GRU4Rec (random) & 1.421 & 1.324 & 1.51 & 1.324 & 1.427 & 1.336 & 1.498 & 1.336 & 1.475 & 1.326 & 1.521 & 1.326 \\
\midrule
SASRec (educated) & N.A. & 1.0 & N.A. & 1.21 & N.A. & 1.0 & N.A. & 1.334 & N.A. & 1.0 & N.A. & 1.326 \\
BERT4Rec (educated) & N.A. & 1.0 & N.A. & 1.341 & N.A. & 1.0 & N.A. & 1.338 & N.A. & 1.0 & N.A. & 1.339 \\
GRU4Rec (educated) & N.A. & 1.0 & N.A. & 1.324 & N.A. & 1.0 & N.A. & 1.335 & N.A. & 1.0 & N.A. & 1.326 \\
\bottomrule
\end{tabular}
}
\caption{Hamming Distance Metric: We report the Hamming distance values for all cases, where smaller values indicate better performance. un\_un, targ\_un, un\_cat, targ\_cat indicates respectively Untargeted-Uncategorized, Targeted-Uncategorized, Untargeted-Categorized and Targeted-Categorized respectively. N.A. denotes instances where the baseline method do not apply in that scenario.}
\label{tab:edit_distance}
\end{table*}

In this section, we present our results. As discussed in \Cref{sec:backgroundandprel,sec:problem_complexity}, we have already addressed \textbf{RQ1}, demonstrating that the problem of finding counterfactuals for a sequence is NP-Complete. Through our results, we aim to answer the following two research questions:
\begin{itemize}
    \item \textbf{RQ2}: How does enforcing counterfactual constraints on a specific target item impact recommendation outcomes?
    \item \textbf{RQ3}: How does generating counterfactuals at the category level, rather than the item level, influence interpretability and effectiveness?
\end{itemize}

\subsection{RQ2: Target Item}

We modified the original sequence using GECE and baseline methods, excluding item targets and categories. The results, shown in \Cref{fig:un_un}, confirm previous findings \cite{betello_ecir_2024,oh2022rank}, demonstrating that even minor changes in the final part of the sequence can significantly impact the model’s output. On the other hand, \Cref{fig:tag_un} illustrates the scenario where targets are considered without categories. In this case, recommending a specific item becomes considerably more challenging due to the large size of the item set $\mathcal{I}$. Moreover, \Cref{tab:edit_distance} reports the edit distance required to reach the desired outcome. Our method achieves often a substantially lower edit distance compared to the baselines, indicating a more efficient transformation process. In cases where the edit distance is comparable or higher with respect to the baselines, our approach consistently yields higher fidelity, highlighting its effectiveness in preserving recommendation relevance.

Furthermore, none of these scenarios have statistical differences in Fidelity with respect to the baselines going to emphasize the difficulty of this problem. Given these observations, we have shifted our focus to category-based recommendations. Identifying a target category presents a more granular challenge, and our findings in this direction will be discussed in the next subsection.

\subsection{RQ3: Target Category}

\Cref{fig:un_cat} illustrate the untargeted categorized scenario: in this case the item set $\mathcal{I}$ is the number of total categories within the dataset. Our method consistently outperforms the baselines: across all three datasets and three models considered, our approach achieves a model fidelity@1 of one, indicating perfect alignment with the original model's predictions. As $k$ increases, fidelity gradually decreases, following a natural trend. While the baselines exhibit a similar pattern, they consistently underperform compared to our method.

Additionally, \Cref{fig:tag_cat} present results for the targeted categorized setting, where the objective is to identify a specific target category. In this scenario, all baseline methods perform significantly worse than GECE, further demonstrating the effectiveness of our approach.

As previously mentioned, \Cref{tab:edit_distance} presents the edit distances required to reach the target outcomes. Our method often outperforms the baselines by requiring significantly fewer modifications, indicating a more efficient and effective transformation process.

Statistical analysis revealed that our method achieved significantly higher Fidelity scores compared to all baseline approaches ($p < 0.01$) in these scenarios, indicating robust and consistent performance advantages.

These results reinforce the robustness and efficacy of our method, highlighting its superiority in both untargeted and targeted categorized settings.

\section{Conclusion}
This paper introduced the problem of generating counterfactuals for sequential recommender systems. First, we demonstrated that this problem is NP-complete by reducing it from Vertex Cover (Theorem~\ref{thm:np}). We then proposed GECE (GEnetic Counterfactual Explanations), the first counterfactual method designed explicitly for sequential recommendations. GECE leverages a genetic algorithm to generate a counterfactual sequence $S'_{u_l}$ that closely resembles the original sequence $S_{u_l}$  while modifying it meaningfully.

We conducted experiments on three diverse datasets to evaluate our approach and tested it across three different models, ensuring robustness and generalizability. We conducted experiments across four distinct contexts, varying in whether counterfactual recommendations are targeted at a specific item and whether they consider item categories rather than individual items. 
Our results demonstrate that GECE effectively outperforms baseline methods in model fidelity and edit distance, highlighting its capability to generate meaningful and efficient counterfactuals. These findings underscore the importance of developing more advanced counterfactual generation techniques for sequential recommender systems. 

To the best of our knowledge, there are currently no counterfactual approaches designed explicitly for SR settings. This highlights the novelty of our proposed method and underscores its contribution in a largely unexplored area. While existing general-purpose counterfactual techniques offer valuable insights, adapting them effectively to SR tasks remains a complex challenge. This work represents an important step in this direction, laying the groundwork for future research to further refine and extend counterfactual reasoning within the SR context.




\bibliographystyle{ACM-Reference-Format}
\bibliography{biblio}

\end{document}